# Interface Framework for Human-AI Collaboration within Intelligent User Interface Ecosystems

Shruthi Andru

Adobe Design, shruthia@adobe.com

Shrut Kirti Saksena

Adobe Design, ssaksena@adobe.com

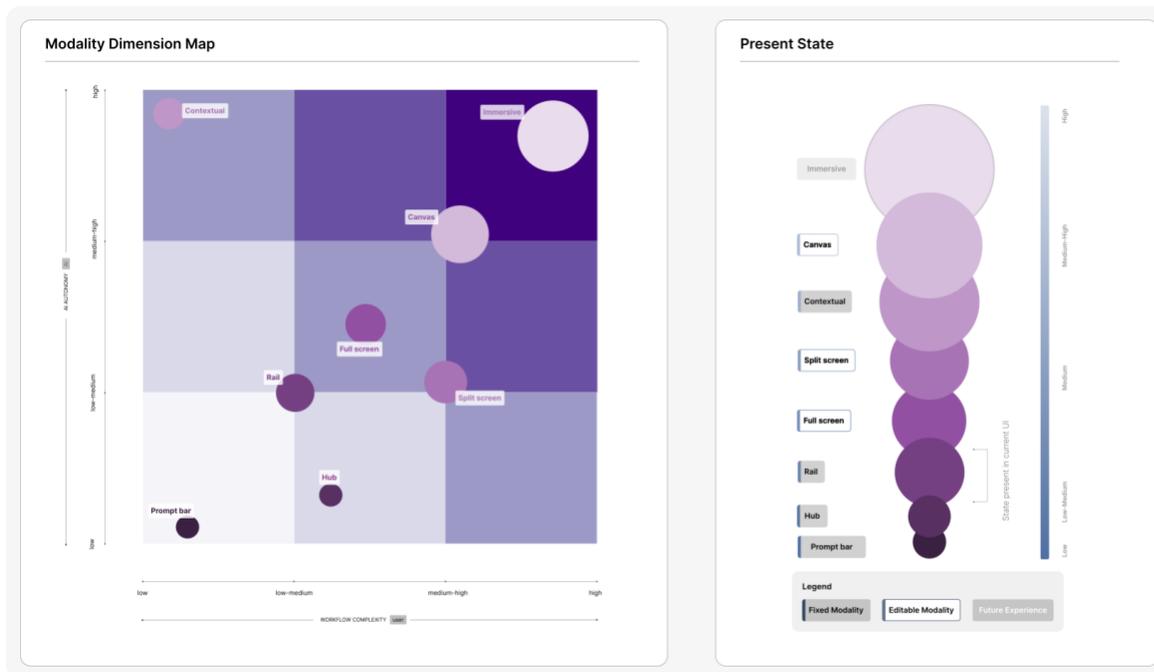

Figure 1: A three-dimensional framework consisting of workflow complexity, AI autonomy, and AI reasoning providing designers with a structured method to assess and select an appropriate interface modality(s) based on specific task types and AI capabilities. We propose a three-step approach: first evaluating the context based on risk; second a set of dimensions and finally selecting the most suitable modality per the evaluation.

As interfaces evolve from static user pathways to dynamic human-AI collaboration, no standard methods exist for selecting appropriate interface patterns based on user needs and task complexity. Existing frameworks only provide guiding principles for designing AI agent capabilities. We propose a dimensional framework based on workflow complexity, AI autonomy, and AI reasoning to guide the design of context-aware, scalable AI interfaces aka modalities[1] (e.g., prompt bars, split screens, full screens, etc.). The framework was developed through co-design workshops with designers of marketing products and refined through qualitative research with eight long-term AI users. The study evaluated the three dimensions, identified task-to-interface relationships, and surfaced the importance of both business impact and security risk across all high-autonomy scenarios. This framework provides product teams with a shared language to develop scalable AI interfaces, emphasizing fluidity between interfaces and progressive user control to balance AI autonomy with human oversight.

## 1  INTRODUCTION

Interfaces form a necessary affordance in today's software applications, bridging legacy SaaS tools with autonomous systems of the future [2,21]. Interactions within these interfaces act as a medium through which humans and agents collaborate to share intent and understanding of tasks that need to be accomplished. However, traditional interface design is rooted in static workflows and predetermined user pathways. This proves to be inadequate for the dynamic, intent-driven interactions that human-AI and multi-agent collaborations in today's advanced applications need.

Rapid AI application development cycles across enterprises, driven by competitive market pressure, have outpaced the development of systematic design methodologies for human-AI (H-AI) interface patterns. This leads to a rigid integration of AI capabilities into existing, disconnected SaaS ecosystems, adding to an already fragmented user experience. They also hinder the deployment of scalable AI solutions. Design decisions regarding AI are often centered on model capability and interaction types [9], rather than the form through which these capabilities manifest.

Furthermore, these AI interactions are often designed as isolated UI components without considering the broader context of user workflows. Recent research has highlighted the importance of contextual factors in human-AI interaction, with factors such as task complexity, AI reasoning, and AI autonomy significantly affecting the effectiveness of human decision-making and collaboration with AI [13,18,31,38,57]. However, these insights have not been synthesized into an actionable framework that product teams can apply, leading designers to make UI decisions based on intuition rather than a systematic evaluation of human-AI collaboration principles.

To address this problem and support this new reality, we propose a framework as shown in Figure 1. to guide the design of composable and scalable AI experiences [16]. Here, interfaces are positioned as first-class entities governed by contextual factors [47]. The framework enables designers to select an appropriate AI interface pattern based on workflow complexity, AI autonomy, and AI reasoning needs. It can be broken down into three parts:

1. A taxonomy of eight AI interface modalities that capture the type of H-AI interactions needed for effective human-AI collaboration in enterprise contexts.
2. A three-dimensional evaluation system based on workflow complexity, AI autonomy, and AI reasoning.
3. An evaluation method that includes risk assessment, dimensional evaluation, and interface pattern mapping to guide cross-functional product teams to make informed design decisions.

The work in this paper also aims to:

---

[1] Throughout this paper, modalities are referred to as interface patterns that we discuss in detail in section 3.1.1.

1. Establish use cases and detailed characterization of each of the eight modalities.
2. Identify and map relationships between workflow complexity, AI autonomy, and AI reasoning levels for each modality.
3. Provide an understanding of progressive user control to balance AI autonomy with human oversight.

The framework was developed through a Research-Through-Design approach [54], consisting of three structured co-design workshops with twelve designers, which emphasized collaborative knowledge and shared decision-making. This was then followed by qualitative interviews with eight practitioners from the field to evaluate the framework.

The implications of this work extend beyond interface design to add to the existing principles of effective human-AI collaboration. This framework has the potential to improve H-AI collaboration outcomes through increased efficiency in decision-making, reduce design inconsistencies, and provide modular patterns that accelerate the development of AI enterprise experiences. As AI capabilities continue to advance, frameworks like this ensure that interface design leverages emerging functionalities while remaining grounded in human-centered design principles.

## 2 RELATED WORK

### 2.1 Human-AI Collaboration

Human-AI Collaboration (HAIC) has emerged as a critical area of HCI research and is defined as the coordinated pursuit of shared goals [9,11,15,17,19,22,23,30,50]. Designing effective collaboration depends on H-AI interaction, adaptability, explainability and trust [5,6,13,18,19,31,32,44,45,48], leading to an increase in human capacity for decision making and problem solving [25,37,46].

Considerations for designing effective HAIC were previously identified, ranging from intelligibility to agency [19,27,49]. Frameworks like ADEPTS [14] also provide agent capabilities, such as disambiguation and transparency, but fail to demonstrate how these dimensions can be effectively used to design an user-facing interface.

### 2.2 Interaction Layers in HAIC & Need for Interfaces

The interaction layer serves as the primary collaboration surface for users to express their intent and interpret system behavior, as proposed by several frameworks for effective HAIC [14,15,31,33,34,36,40,50,51]. One of the main challenges in designing interactions for evolving adaptive systems is the lack of predictable outputs. Moreover, mental models play a key role in AI-assisted human decision-making [4] and are critical to creating effective collaboration. This highlights the need for consistent and fluid interface patterns that facilitate both simplified decision-making and effective human-AI collaboration [20,53].

### 2.3 Beyond Fixed UI Silos

A survey of current AI systems shows that most AI interaction patterns are confined to fixed approaches where human interactions with the interface mainly consist of basic actions like menu selections or button clicks [24,26]. This suggests a broader pattern of AI interfaces being designed as fixed, standalone components rather than adaptive, integrated systems. This forces humans to bridge gaps across these disconnected solutions instead of having a seamless handoff experience across agents, interfaces, and tasks. An example of this fragmentation is evident in conversational systems that are often considered to have isolated, stateless interactions that lack integration with broader workflows [24,41,52,56].

## 2.4 Current Frameworks and their Limitations

Several frameworks [14,33,34,40,51,31,15,45] have been proposed to guide the design of human-AI collaboration, each stating principles for how humans interact with agents, but fail to describe how UI elements of the interaction layer work together [10,12,50]. While these frameworks outline detailed thinking on collaboration, they do not provide practical methods for making design decisions about selecting appropriate patterns for specific contexts. ADEPTS provides comprehensive guidance on agent capabilities but fails to provide a systematic method for determining how and where users should interact with these capabilities. While the field of HCI has developed a robust understanding of H-AI collaboration principles and evaluation methods, the limitations of existing frameworks highlight a gap in a standardized, actionable method for selecting appropriate interface patterns based on contextual factors, such as task complexity.

## 3   THE FRAMEWORK

We propose an AI interface evaluation framework developed via a Research-through-Design approach [54] that provides designers, researchers, and developers with a structured method to design & develop intelligent user interfaces (IUI) for HAIC in complex enterprise workflows. The three-dimensional framework enables the systematic selection and design of AI interface patterns (modalities) through a structured evaluation of users' workflow complexity, the degree of AI reasoning that an interface can support, & AI autonomy needed to accomplish a task. At its core, the framework aims to transition AI solutions designed for HAIC from isolated and fragmented experiences to composable, scalable patterns [16]. Additionally, the framework is domain-agnostic and can be applied to any complex enterprise ecosystem that requires human-AI collaboration.

### 3.1   Core Components of the Framework

The framework consists of two main components: modalities and dimensions.

#### 3.1.1 Modalities

In this paper, we refer to modalities as an interface pattern or form that manifests how the AI's presence and capabilities are presented to the users. A modality functions beyond a UI container; each modality embodies design decisions about how visible, assistive, proactive, and agentic the AI is for a specific task the user intends to accomplish. The framework proposes eight distinct modalities describing the form AI interfaces should take, and the three dimensions providing the selection criteria for choosing appropriate modalities based on the task. As shown in Figure 2., the modalities are divided into four categories based on interaction, collaboration, and contextual requirements that the user has.

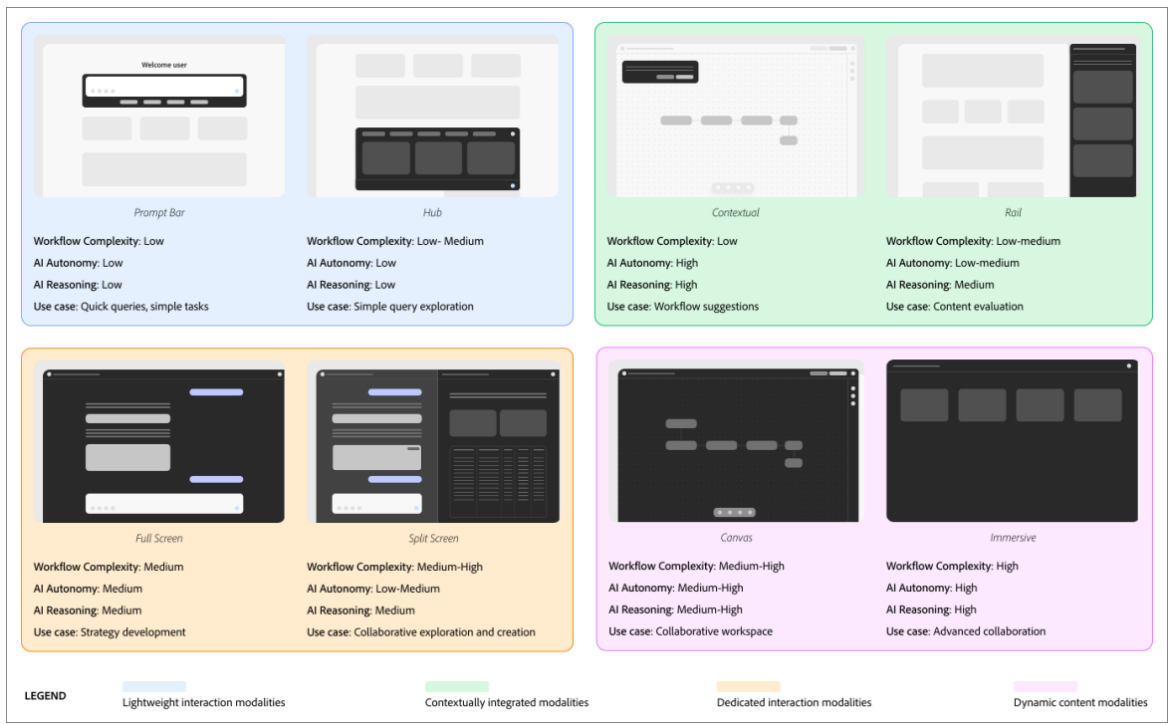

Figure 2: Eight modalities based on the three dimensions in the framework

Lightweight Interaction Modalities: These modalities facilitate quick and focused interaction with minimal cognitive load for the user. They are designed for rapid task completion without disrupting primary workflows and are information retrieval modalities[2]. They consist of:

1. **Prompt Bar:** The prompt bar offers users a compact, search-like interface with accompanying action buttons for quick actions. The interface appears as a horizontal input field where users can type in natural language. This modality is best suited for specific, single-step information queries.

2. **Hub:** The hub is an expanded version of the prompt bar, offering users a command center-like interface to view recent interactions or suggested actions. This modality surfaces rich contextual data, an integrated chat interface and the ability to search and review objects. The hub also serves as a launching point for AI interactions and is best suited for users who need to explore AI capabilities or initiate complex workflows without needing a dedicated space.

---

[2] Information Retrieval Modalities are lightweight entry points designed for fast information retrieval and low-stakes queries. These allow users to quickly surface facts, run simple natural language commands, or retrieve insights with minimal friction.

Contextually Integrated Modalities: These modalities embed AI assistance directly within existing workflows, allowing for seamless AI integration. They are designed to provide intelligent support that seamlessly integrates within the user's current workflow and considered to be companion modalities[3]. They consist of:

3. **Contextual:** The contextual modality provides users with AI-driven suggestions, insights, or nudges by surfacing potential edits or improvements they can make, all within the context of their workflow. They appear to users as inline suggestions, smart annotations, or contextual tooltips without disrupting the user's primary working area. This modality is best suited for proactive, transparent, actionable recommendations that users can engage or dismiss as needed.

4. **Rail or Panel:** The rail modality provides users with a persistent AI presence in the context of their workflow through a docked side panel interface without obstructing the primary working area. This modality appears on the right side of the screen, containing conversational elements, controls, and additional contextual information related to the user's current task. The rail is best suited when users need both reactive and proactive AI assistance, i.e., initiate conversations as well as receive contextual guidance. The persistent nature of the rail enables it to be present across applications while maintaining context and conversational history as users navigate through multi-application workflows.

Dedicated Interaction Modalities: These modalities provide focused environments designed for intensive H-AI collaboration. Designed to offer a dedicated space that prioritizes AI interaction over other interface elements. They consist of:

5. **Full screen:** The full-screen modality provides users with a full-page, immersive environment to interact with the AI for deep-focus tasks over an extended period. The interface features a large conversation area, dynamic cards[4] with rich media display and controls for managing complex AI interactions while preserving conversational history. The modality is best suited for supporting multi-turn conversations and tasks, such as detailed analysis, content generation, and ideation, that benefit from having maximum screen real estate.

6. **Split-screen:** The split-screen modality offers users an adjustable, divided interface that provides equal focus on both AI assistance and primary content. The left side of this modality features the AI interaction interface, which includes conversation threads with dynamic cards, and the right portion displays primary content, such as documents, data visualizations, and the application surface. This modality is best suited for users who require real-time collaboration with AI, enabling them to rapidly iterate on content while receiving immediate AI feedback and hence is referred to as an action-oriented modality[5].

Dynamic Content Modalities: These modalities provide users with a space for both user-generated and AI-generated artifacts. They are designed to provide users with the flexibility to manage, manipulate, and interact with AI-generated content in their workflows. They consist of:

7. **Canvas:** The canvas modality provides users with a flexible workspace where they can develop AI-generated content by directly manipulating the interface. This modality features an open work area with drag-and-drop functionality, resizable content blocks, and supports spatial reorganization based on workflow requirements. The modality is best suited for users who require version control capabilities, complex editing tools for manipulating visual layout, and collaborative features that enable both individual and team-based creative processes.

---

[3] Companion Modalities stay alongside the workflow to provide context, guidance, or augmentation without taking over the task. They feel like a partner or "sidekick."

[4] Dynamic cards are recognizable, repeatable patterns for information that AI returns to the user in any modality.

[5] Action-oriented Modalities where the primary task work happens and either users or AI directly manipulate, edit, or create in these spaces. These carry the weight of execution by returning an output for every input provided.

8. **Immersive:** This modality provides users with an advanced collaborative workspace where the AI transcends the role of a traditional AI assistant to become an active co-creator. The modality anticipates fluid interaction patterns where humans and AI can initiate actions, modify shared artifacts and engage in deep problem solving across workflow types. While current AI capabilities and interface technologies limit immediate design and implementation, including this modality in the framework ensures that future systems are built keeping this foundation in mind [6,23,46,56].

*3.1.2Dimensions: criteria for evaluation of modalities*

Dimensions are qualitative levers that enable product teams to select the most suitable modality based on the context of the interaction needed for effective HAIC. This ensures that AI presence and behavior are consistent, creating fluid and intuitive experiences across diverse interaction scenarios and collaboration needs. There are three main dimensions:

1. Workflow complexity: This is a user-centered dimension that refers to the overall complexity or demand of the user's task [10,12,31]. This dimension helps product teams determine how much support or guidance AI should provide for simple one-step tasks to dynamic multi-step workflows that require more proactive assistance. There are four levels of complexity, ranging from low to low-medium, medium-high, and high. It is influenced by four equally weighted factors:
   - Number of steps involved
   - Dependency on other tasks or objects in the platform ecosystem
   - Experience of the user, i.e., beginner vs expert
   - Coordination needed across tools, users, or data sources.

2. AI autonomy: This dimension defines the degree of independence AI systems have while acting on behalf of the users. It helps product teams strike a balance between user control and the extent to which AI drives the experience [33,43,49]. The levels of AI autonomy range from low, where the AI waits for explicit user instructions, to high autonomy, where the AI proactively makes decisions and executes complex multi-step tasks with minimal user intervention [7,12,33,33,38,39,44].

3. AI reasoning: This dimension refers to the AI's ability to understand context and solve problems [27,31]. It helps teams determine how much intelligence is needed to support workflows and user experiences. In this dimension, we prescribe three levels of reasoning on a spectrum from low to high, as shown in Figure 1.

## 3.2 Guardrails and Risk Mitigation

All areas involving high AI autonomy should be designed with careful consideration, incorporating design guardrails that provide both transparency and explainability of AI actions as well as user control mechanisms [13,27,32,45]. This ensures that users can intervene and redirect AI actions at any point in the interaction thereby maintaining human authority over AI systems.

There are two primary categories of potential impact when assessing a task. First is compliance and security risk, which assesses whether a task or workflow involves regulated data or compliance requirements, such as HIPAA or GDPR, in regulated industries like healthcare or financial domains. Second, is business impact assessment, which focuses on the consequences of incorrect task execution that causes potential brand perception implications if the AI system makes errors or is compromised.

### 3.3    Framework Operation and Modality Selection

#### 3.3.1 Present and Future State Considerations

AI capabilities and technology are advancing at a rapid pace [58]. Current enterprise AI developments primarily operate between low and medium reasoning capabilities, where Prompt bar, Hub, and Contextual modalities offer lightweight interactions, while Full-screen and Split-screen modalities support more advanced interactions. As AI evolves towards multi-agent collaboration and higher autonomy, we anticipate that future state modalities, such as Immersive and Canvas, will support deeper reasoning capabilities.

#### 3.3.2 Fluid Transition Mechanism

The framework prioritizes user control in selecting modality entry points[6], and transitions between these modalities should be designed to reflect the same. Users should be able to initiate interactions from any of the given modalities and seamlessly switch to another if needed, based on control, visibility, or space [53].

#### 3.3.3 Modality Selection

The modality selection process involves a systematic three-step process for selecting the appropriate AI interface modality based on risk evaluation and dimension assessment, as shown in Figure 3.

1. Risk Assessment: First, product teams evaluate both categories of risk, namely compliance and security risk, as well as business impact risk. High-risk tasks require guardrails to be implemented before designing AI solutions. Low- to medium-risk tasks can continue to assess dimensions and select interface modalities.
2. Dimension Assessment: The next step in this process is to evaluate the dimensions in the following sequence. First, we assess workflow complexity based on the number of steps, the coordination required across users, tools, and data sources, and the user expertise levels. Next, we determine the level of AI independence based on task characteristics, users' tolerance for risk, and comfort with AI making decisions.
3. Modality Mapping: Teams then use the modality graph as shown in Figure 3., to identify the best fit for their use case.

---

[6] Modality entry points are defined as controls within an interface that are either UI or natural language and let users transition from one modality to the next.

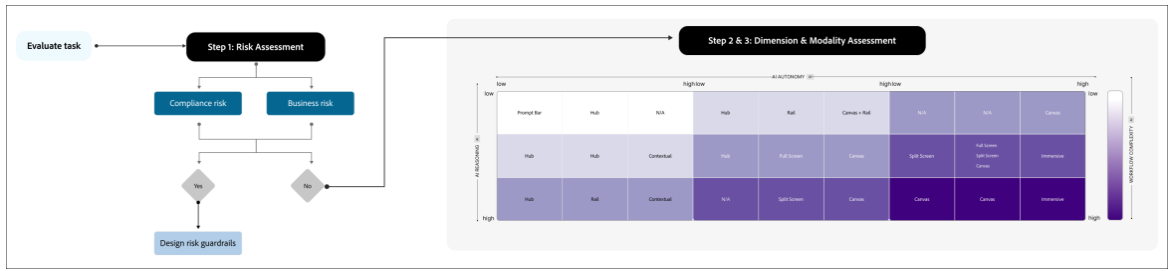

Figure 3: Flow-chart showing the three-step process based on risk evaluation and dimension assessment

## 4 METHODOLOGY

Using a Research-through-Design approach [54], we developed this framework in two distinct phases. First, we ran three structured participatory co-design workshops with designers of marketing products to collaboratively construct the framework [42]. Second, we conducted semi-structured interviews with marketing practitioners in the field who had used AI tools for more than 5 months and evaluated the framework through a think-aloud protocol using an AI-generated prototype that employed these modalities.

### 4.1 Co-Design workshops

The Co-Design process was structured into three 90-minute workshops, conducted over a three-week period. Each session employed specific participatory activities, allowing for progressive development and iteration of the framework. This approach enabled us to elicit tacit designer knowledge and collectively create an artifact that was developed through shared decision making [35,41,42,55,59].

#### 4.1.1 Participants

The workshop engaged twelve UX Designers from multiple product teams within a large enterprise ecosystem, spanning customer data, analytics, content management, and journey orchestration domains. All participants had a robust understanding of AI interfaces and had previously designed AI solutions for marketing products. The workshop was facilitated by two lead designers who had also previously designed enterprise AI and AI agent solutions.

#### 4.1.2 Procedure and Data Collection

The three sessions are as follows:

1. Session 1: Modality alignment and brainstorming dimensions

The primary objective of this first session was to establish a set of modalities based on existing AI interface patterns and to begin brainstorming a set of dimensions that could be used to determine when each modality should be applied. The first 15 minutes of the session were spent identifying existing terms and AI interface patterns in the industry. The next 25 minutes were spent individually listing definitions of these modalities and how they could be used in the given product context, using sticky notes. Subsequently, for the next 35 minutes, the participants engaged in a prioritization exercise using dot voting, followed by an open discussion to reach a consensus on the definition and use case of each modality. The remaining 15 minutes of the session were spent brainstorming metrics or dimensions used to evaluate AI UX [28,29]. The outcome of this session was the selection of seven main modalities, namely: Prompt bar, Hub, Preview, Full screen, Split screen, and Canvas.

2. Session 2: Dimension alignment

The second session focused on aligning on a set of dimensions that enabled designers to evaluate how a modality should be used. Designers spent the first 10 minutes revisiting the work from the previous session. A significant portion of the session, 25 minutes, was dedicated to clarifying the terminology of each dimension listed and resolving conceptual overlaps using sticky notes, dot voting, and moderated discussion. The session then continued with another activity, using the same methods to define the levels, impact, and boundaries of each dimension.

The session concluded by identifying the three main dimensions of the framework: workflow complexity (WC), AI autonomy (AIA), and AI reasoning (AIR), which are divided into two categories: user considerations and AI controls.

3. Session 3: Determine workflow-to-modality and modality-to-dimension mapping

The final co-design session was dedicated to mapping existing user workflows to modalities and mapping AI modalities across the dimensional axes. For the first 30 minutes, designers utilized the existing Jobs-To-Be-Done map [60] for each product and mapped them to the seven modalities. Next, in a 15-minute activity, the designers mapped the modalities on a scale of low to high for AI reasoning through a facilitated discussion. The final 45 minutes were spent plotting the modalities onto a two-dimensional map, with user complexity and AI autonomy ranging from low to high, considering the level of AI reasoning each modality was capable of. This led to the identification of an additional, future-looking modality, Immersive, that encompasses all the other seven modalities.

The key outcome of this session was a consolidated framework featuring eight core modalities, a set of three dimensions, identification of areas of risk that need design guardrails, and planned to understand how modalities could transition into each other through semi-structured user interviews.

*4.1.3Analysis*

Following collaborative analysis principles [35] designers engaged in a collective interpretation of workshop outputs, rather than relying on a solo designer or researcher-only analysis. The democratic consensus-building enabled transparent and consolidated decisions, leading to the development of a framework that was aligned across differing product domains.

## 4.2 Semi-Structured Interviews

We evaluated the framework through 90-minute semi-structured interviews and think-aloud protocols [20] with eight marketing professionals who were long-term AI users, utilizing an interactive AI-generated prototype for the sessions [61]. This approach enabled us to gather feedback on the effectiveness of modality across different task complexities and understand user preferences regarding AI reasoning and autonomy controls.

*4.2.1Participant Recruitment and Demographics*

The participants were recruited using the enterprise company's customer database through a screener questionnaire [3]. The primary inclusion criteria required participants to be:

1. Actively engaged in marketing activities such as demand generation, campaign execution, performance analysis, marketing automation, nurture campaign management, audience segmentation, audience activation data management, and cleansing.

2. Utilize marketing tools to support these activities at least three to four times a week.
3. Have experience experimenting with AI tools, including an AI assistant within their workflows.

Additionally, participants were required to review and sign informed consent agreements before sharing Microsoft Teams link to join the session.

The study consisted of eight participants, all of whom were above the age of 18, from the US and Canada. Based on established practices in qualitative evaluative studies, the resulting distribution provided us with participants who engaged in a wide range of marketing activities and were long-term AI users who would benefit from the systematic AI interface designs proposed by the framework.

### 4.2.2 AI generated prototype

We used an interactive prototype built using Cursor[7] to understand how users transition across modalities. Users begin with the Prompt bar on the home page using natural language input, which then transitions them to the Full screen for complex prompts like "create an audience". The AI generates step-by-step plans with actionable steps. Selecting audience creation presents the user with a dynamic card that opens in the split-screen modality. The second step involves journey creation, producing a dynamic card with a button to "Open journey editor" that transitions them into a canvas-rail hybrid interface. The canvas provides them with a static workspace with suggestions from the contextual modality pre-populated, and the rail offers a docked conversational interface with transition controls to expand to the full-screen modality. Additionally, the floating action bar triggers the rail from the canvas, and a home button in the side navigation allows users to return to the starting point of this workflow.

### 4.2.3 Procedure and data collection

The remote 90-minute sessions consisted of three phases:

1. Phase 1: The sessions began with the introduction of the study's objectives, procedures, and expectations, and participants were asked to self-identify their level of usage of AI tools. Participants then described their role, company context, day-to-day responsibilities, marketing tool usage patterns, and experience with AI experimentation.
2. Phase 2: This was followed by a think-aloud testing protocol with the interactive prototype. The participants were asked to reflect on the prototype within the context of audience and journey management workflows. They were specifically measured on the intuitiveness of the modality elements, usability of the experience, expectations for varying levels of task complexity, preferences around user control versus AI autonomy, and expectations for the level of AI reasoning supported by each modality.
3. Phase 3: The last phase concluded with the participant's reflection on their preferences, perceived benefits, concerns, and desired future interactions with AI modalities.

### 4.2.4 Analysis

Interview transcripts were analyzed using a thematic coding approach [1,8], identifying themes and insights related to the effectiveness of the modality framework. Key themes included modality suitability across varying task complexities, preferences around AI autonomy and user control, expectations for AI reasoning, and the seamlessness of transitions between modalities.

---

[7] Cursor is a vibe-coding tool that uses generative AI to develop interactive prototypes.

# 5  CASE STUDY

To demonstrate the practical application of our framework, we present two marketing use cases that illustrate how the three-dimensional evaluation system guides the selection of modalities. Each case study provides details of the user's context, marketing task requirements, modality selection rationale via the dimension assessment, and task execution details. The modalities were chosen by assessing the sub-tasks within each task and carrying out the modality selection assessment, as shown in Figure 4. and Figure 5., on each of these subtasks, determining the level of dimension for each task type.

These case studies are low-risk real use cases from existing products that demonstrate how the framework enables designing contextually appropriate interfaces for similar workflow types across different products. For the case study, all the tools used were from a single suite of cloud-based marketing platform tools.

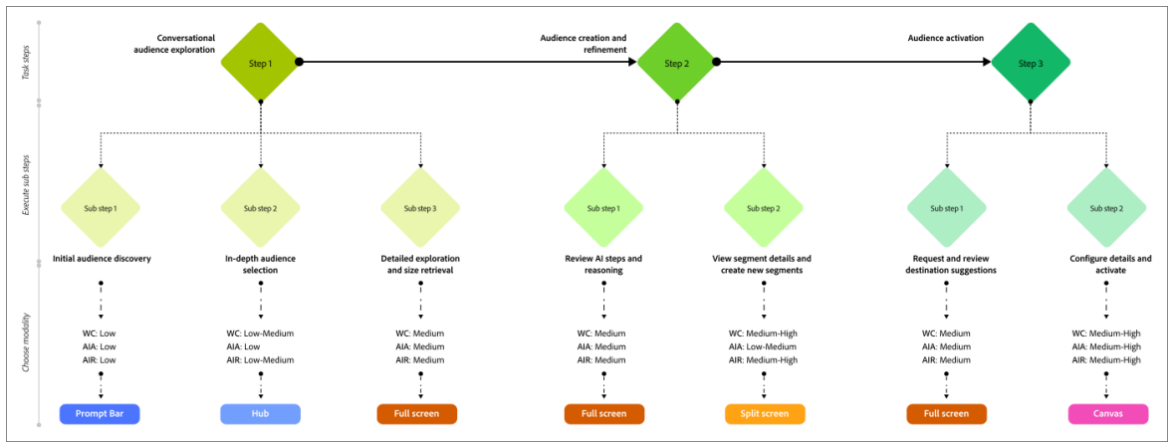

Figure 4: Flow-chart showing modality selection for creating, editing, and activating audiences in cloud-based platform

## 5.1  Case study 1: Creating, editing, and activating audiences[8] in cloud-based platform for collecting and analyzing customer data

### 5.1.1 User context

**User:** Brand media manager at a global footwear company

**Role:** Responsible for audience targeting for digital campaigns

**Goal:** Launch a new campaign promoting eco-friendly running shoes across North America

### 5.1.2 Workflow analysis and modality selection

1. Step 1: Conversation audience exploration

**Sub-tasks:** There are three sub-steps, namely, initial audience discovery, in-depth audience selection, detailed exploration, and audience size retrieval

---

[8] An audience is a group of people that a system or campaign is designed to reach

**Task execution:** The user begins with the prompt bar modality to initiate the initial query for finding audiences for the campaign. The user then switches to the hub modality to access more in-depth audience selection without leaving the home page. They then transition to the full-screen modality for more detailed exploration, such as retrieving audience sizes, etc.

2. Step 2: Audience creation and refinement

**Sub-tasks:** There are two sub-steps, namely, review AI steps and reasoning, view segment[9] details, and create new segments

**Task execution:** The user continues the conversation in the full screen to review the steps suggested by the AI. The AI then provides a few previously created segments. The user then transitions to the split-screen to view more details and creates a new segment from the suggestions. This also provides the user with the ability to refine the audience while maintaining visibility of the AI insights.

3. Step 3: Audience activation[10]

**Sub-tasks:** There are two sub-steps, namely, request & review destination[11] suggestions, and configure details & activate

**Task execution:** The user switches back to the full screen modality and asks the AI to suggest destinations to activate this audience. After reviewing and selecting the destinations, the user chooses to open the audience in the canvas to configure additional details and activate.

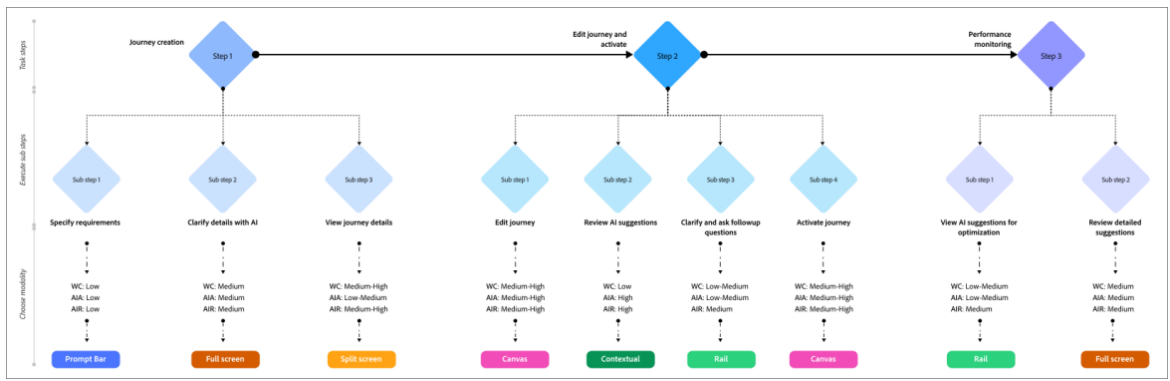

Figure 5: Flow-chart showing modality selection for journey creation and monitoring in cloud-based platform for campaign execution

## 5.2 Case study 2: Journey[12] creation and monitoring in cloud-based platform for campaign execution

### 5.2.1 User context

**User:** Marketing automation specialist

**Role:** Responsible for designing and optimizing journey campaigns

**Goal:** Create a marketing journey to increase premium subscriptions for their customers

---

[9] Segments are smaller sub-groups within an audience grouped by shared characteristics like behavior, demographics or preferences
[10] Audience activation refers to the process of engaging a chosen segment group by delivering information, offers or messages through digital marketing channels.
[11] Destinations are end channels where messages are sent so that experiences can be delivered to customers
[12] A journey is a planned sequence of interactions that the customer experiences across different touchpoints within a marketing campaign.

*5.2.2 Workflow analysis and Modality Selection*

1. Step 1: Journey creation request

**Sub-tasks:** There are three sub-steps, namely, specify requirements, clarify details with AI, and view journey details

**Task execution:** The user uses the prompt bar to convey the request to AI, specifying initial journey requirements. They are then directed to the full screen modality to respond to AI's clarifying questions and is then provided with an AI-generated journey returned in a dynamic card. The user views more details and a preview of the journey in the split-screen. This allows the user to view details while also allowing them to ask the AI follow-up questions about the journey.

2. Step 2: Edit the suggested journey and activate[13]

**Sub-tasks:** There are four sub-steps, namely, edit journey, review AI suggestions, ask follow-up questions, and activate journey

**Task execution:** The user opens the journey in the canvas to view details of each node. This enables the user to view all the details and adjust the node placements as needed. The AI then suggests changes via the contextual modality, which the user accepts. The user also asks follow-up questions, using the rail modality. The user then activates the journey by selecting activate from the canvas.

3. Step 3: Performance monitoring and optimization[14]

**Task:** There are two sub-steps, namely, view suggestions for optimization and review suggestion

**Task execution:** The user receives an alert and opens the rail to view AI suggestions for new user segments. After reviewing details in the full screen, the user dismisses the suggestion.

**5.3 Framework benefits**

The case studies demonstrate how the framework facilitates fluid modality transitions based on evolving task complexity, enabling progressive control as the user progresses from exploration to execution. The contextual appropriateness of the modalities also matches the cognitive and interaction requirements of each workflow step. As users utilize multiple products from the marketing suite of platform tools, this framework enables them to preserve mental models across products, thereby enhancing the user experience within the ecosystem.

**6 DISCUSSION**

The study helped refine the framework and deepened our understanding of how task complexity, AI autonomy, and AI reasoning interact to determine which modality is most effective for a given marketing task. Below, we discuss some of the findings and implications from the work:

**6.1 Interplay of Task Complexity, AI Reasoning, and Autonomy in Modality Preferences**

Our findings reveal two distinct but interconnected patterns: a linear relationship between task complexity and modality preference, and a non-linear relationship between AI reasoning, autonomy, and modality preferences.

---

[13] Journey activation is the action of deploying the planned sequence of steps for the customer within a marketing campaign.

[14] Performance monitoring and optimization is measuring whether an audience or journey is achieving the goals and metrics to adjust the strategy to further improve outcomes.

*6.1.1Task complexity and modality preference.*

Modality preference here scaled linearly with the increasing demands of task complexity. Simpler, low-stakes tasks were expected to be carried out with lightweight modalities, such as the prompt bar, contextual, or rail, which were valued for their speed, ease, and unobtrusive guidance. As task complexity increased, participants gravitated towards more immersive, tangible button-and-click modalities such as split-screen and canvas. These modalities afforded greater control, visibility, reversibility, and collaboration that were deemed essential for managing multi-step, non-linear, and precision tasks.

*6.1.2 AI reasoning and context.*

The expectations around AI reasoning and autonomy diverged from this linear pattern. As shown in Figure 6., reasoning requirements did not simply increase with task complexity but flexed dynamically depending on the context and task complexity that could be handled by the modality. For example, in canvas and split-screen, participants expected high reasoning depth, stepwise explanations, artifact[15] integration, and layered guidance. We found that they were most suitable for complex, non-linear workflows and precision tasks. Conversely, modalities like contextual and rail were expected to provide high levels of intelligent reasoning, surfacing timely clarifications or optimizations without interrupting the flow.

Another notable factor was that as reasoning depth increased, participants expected modalities to provide more space, richer artifacts, and tighter integration, except in cases where reasoning was expected to enhance the flow without disrupting it, such as the contextual modality.

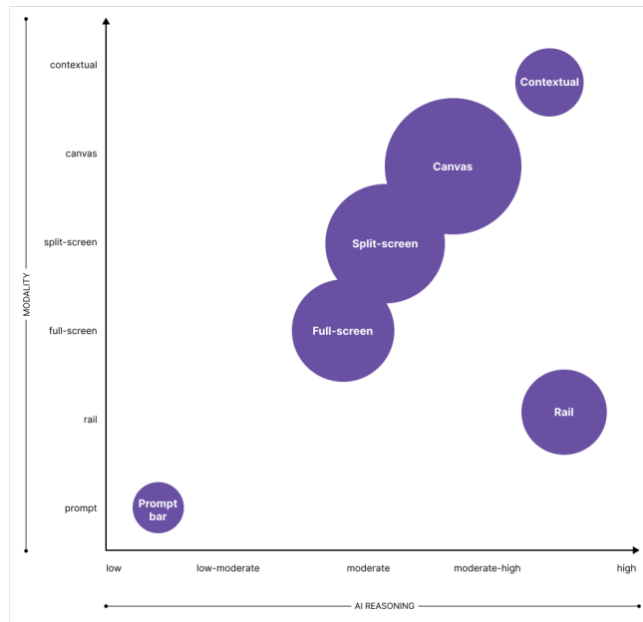

Figure 6: Non-linearity in AI reasoning spectrum within modalities

A central theme that emerged from the study was the importance of how AI reasoning is communicated across modalities. Participants not only expected the AI to explain why a particular response or suggestion was generated but also required that this reasoning be delivered in a way that aligned with the constraints and affordances of the interface, making modality-specific reasoning essential. Mismatches between the explanation format and modality constraints, such as overly verbose outputs in limited spaces, led to frustration and eroded trust in the AI. This finding highlights the importance of information density and the necessity of tailoring explanations to the visual capacity of each modality. Figure 7. illustrates this relationship by mapping information density against task complexity and modality space.

---

[15] An artifact is an AI generated output.

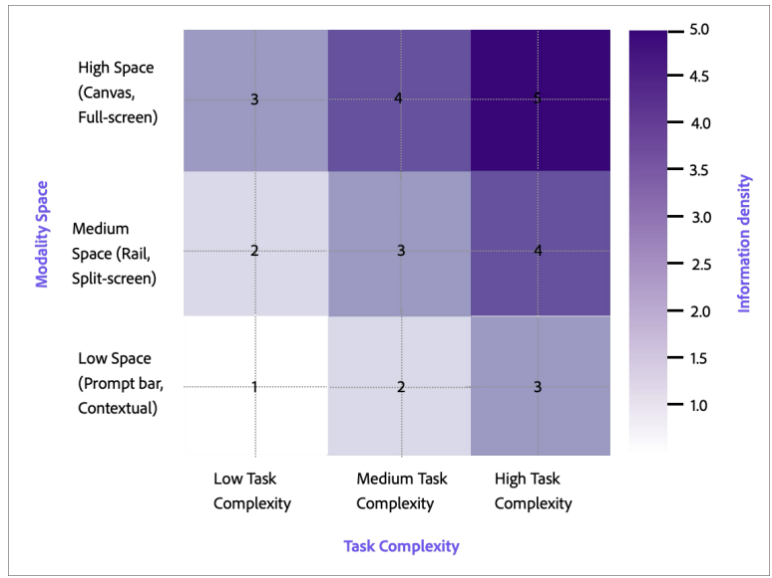

Figure 7: Mapping information density against task complexity and modality space

Thus, effective modality design must ensure that explanations are both context-sensitive and space-appropriate, sustaining trust and accountability without overwhelming or distracting the user.

### 6.1.3 AI autonomy spectrum and modalities

Modality preferences also reflected trade-offs between complexity and autonomy. Modalities capable of handling high complexity, such as split-screen and canvas, did not necessarily support high autonomy, as participants often desired greater control in such contexts. These were expected to balance autonomy with user control for complex workflows and were perceived as human led.

In contrast, modalities such as the prompt bar and rail were seen as AI-led, with the system taking initiative for speed and efficiency in simple or repetitive tasks. These modalities were trusted for automating low-stakes actions. Sitting between these extremes, split screen and full screen were regarded as collaborative modalities, supporting a shared partnership in which both the human and AI contributed to task completion. These modalities were valued for their ability to balance AI autonomy with user oversight, enabling flexible division of labor.

A key design implication is the need for progressive control within AI-led and collaborative modalities. Progressive control ensures that while AI can take initiative and accelerate tasks, users retain the ability to intervene, adjust, and redirect outcomes at any point. Thus, progressive control provides a bridge between task-based and emerging goal-based workflows. Such flexibility is essential not only for everyday efficiency but also for high-stakes tasks where precision, accountability, and trust are paramount.

## 6.2 The Role of Risk in Modality Preference

Risk emerged as a critical factor shaping user preferences for AI modalities. Participants emphasized that risk amplifies the precision required in execution. When tasks carried high risk, whether low or high complexity, participants consistently favored modalities that emphasized UI controls. These were described as familiar, predictable, and reliable, offering users confidence that errors could be quickly identified and corrected. Additionally, modalities that relied heavily on AI autonomy were seen as less appropriate in such contexts, as they did not provide the same level of transparency or user oversight.

Thus, risk is not merely operational but also strategic and reputational. Designers are recommended to pair high-stakes workflows with user-controlled, transparent, and auditable modalities, while reserving highly autonomous modalities for low-stakes contexts, at least until AI reaches accuracy and verifiability on par with human judgment [48].

## 6.3 Fluid and Bi-Directional Modality Transitions

The study emphasized the importance of fluid, bidirectional transitions between modalities. Participants consistently stressed that transitions should not be one-way or require cumbersome workarounds. Instead, every transition needs to be designed to ensure reversibility, efficiency, and minimal disruption to workflow. For example, as shown in Figure 8., users should be able to move from split-screen to canvas and vice a versa seamlessly, without requiring extra steps or reliance on natural language commands. This expectation reflected a broader desire for flexibility as tasks evolved, with users wanting to move fluidly between modalities as complexity or needs shifted. Moreover, it was observed that UI controls remain essential in the near term for switching between modalities over natural language commands.

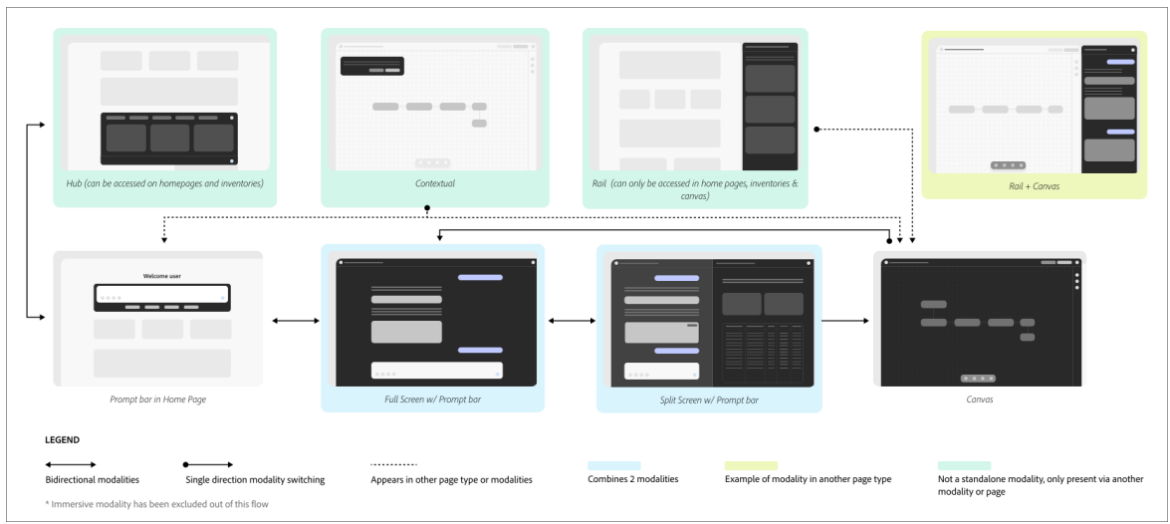

Figure 8: Diagram showcasing fluidity and need for bidirectionality in modality transitions

## 6.4 Implications

### 6.4.1 Extending current understanding of AI Interface design:

Our framework contributes a novel approach to designing cohesive AI experiences across product ecosystems, bridging the gap between theoretical principles and practical implementation.

Prior literature on human–AI collaboration [9,11,15,19,20,22,26,30,31,33,40,50,62] has emphasized autonomy, intelligibility, and task delegation as central considerations in agent-based experiences. Building on this foundation, our work extends the discussion in several important ways. First, we foreground the role of workflow complexity within task delegation, highlighting how the intricacies of multi-step, non-linear workflows shape user needs for both interface design and agent architecture. Second, we introduce information density as a function of both modality space and task complexity, positioning explainability as one component within a broader reasoning process that governs how AI communicates its intent to the user. This perspective advances prior discussions of intelligibility [27] by specifying how explanation [27,32] content tailored to interface constraints to facilitate seamless human-AI collaboration. Finally, while existing literature [50] only addresses workflow complexity at the level of agent architecture, our work contributes a complementary perspective by examining how complexity must also be accommodated in user-facing interface design.

### 6.4.2 Transforming design practice

For design practitioners, this framework provides a new systemic method for making AI interface design decisions. It also enables design teams to work with standardized interaction patterns that help users preserve mental models across different AI-powered applications. This is especially valuable for organizations rapidly developing AI features, where consistent approaches can improve both user experiences and reduce development costs.

The identification of progressive user control addresses concerns about relative AI autonomy and user control and provides design patterns that maintain human oversight along with AI assistance.

The study's findings also suggest that AI systems should be designed with multiple interaction modalities in mind from the beginning, rather than retrofitting interface elements onto existing systems.

## 7 CONCLUSION

Jef Raskin once said, "As far as the customer is concerned, the interface is the product" [63]. Similarly, when it comes to AI powered applications, the interface becomes the sole mediator between human intention and artificial intelligence capabilities. There is more to interface design than simply providing access to AI functionality, it fundamentally shapes how users conceptualize, communicate intent, trust and interact with AI systems. Thus, designing consistent interfaces is the key to Human-AI Collaboration.

In this paper, we mapped use-cases to modalities, established a relationship between three dimensions and provided an understanding of progressive user control to balance AI autonomy with human oversight. The findings show that while modality preferences scale linearly with task complexity, AI reasoning and AI autonomy interact in non-linear, context-sensitive ways to shape effective modality design for marketing tasks. Our proposed interface selection framework bridges the gap between high-level human-AI collaboration principles and design decisions, enabling practitioners to move beyond ad-hoc interface choices to evidence based, contextually appropriate, scalable design solutions.

As the rapid proliferation of AI systems continues, we hope this framework provides designers, researchers and developers with the tools to create effective, scalable interfaces leading to trustworthy and empowering Human-AI collaboration.